\documentclass[prd,showpacs,nofootinbib,preprint,10 pt]{revtex4-2}
\usepackage{amsmath,float,graphicx,color,xcolor,epsfig}
\usepackage{epstopdf}
\usepackage{soul}
\usepackage{hyperref}
\usepackage{slashed}
\usepackage{subcaption}
\usepackage{amsmath,amssymb}
\usepackage{slashed}
\usepackage{graphicx}
\usepackage{xcolor}
\begin{document}
\begin{titlepage}


%
\begin{center}
\setlength {\baselineskip}{0.3in}
{\bf\Large\boldmath
Radiative $B$ to axial-vector meson decays at NLO in Soft-Collinear Effective Theory
}\\[5mm]
\setlength {\baselineskip}{0.2in}
\setlength {\baselineskip}{0.3in}
{\large Arslan Sikandar$^{1}$, M. Jamil Aslam$^{1}$, Ishtiaq Ahmed$^{2}$, Saba Shafaq$^{3}$\\[5mm]
~{\it $^{1}$Department of Physics, Quaid-i-Azam University, Islamabad 45320, Pakistan. \\
$^2$National Centre for Physics, Quaid-i-Azam University, Islamabad, Pakistan.}\\
$^3$ \it Department of Physics, International Islamic University, Islamabad, Pakistan.}\\[5mm]
			%
			
			%
			%
{\bf Abstract}\\[5mm]
\end{center}
\setlength{\baselineskip}{0.2in}
The rare decay $B\rightarrow A\gamma$, with $A$ representing axial-vector mesons such as $K_1 (1270),\; K_1 (1400),\;  b_1(1300),\; a_1(1260)$, is studied at next-to-leading order (NLO) in soft collinear effective theory (SCET). The large outgoing meson energy encourages the study of the decay with an appropriate factorization scheme that separates the factorizable and non-factorizable parts systematically. We have analyzed the leading-power and $\mathcal{O}(\alpha_s)$ diagrams that contribute to matching to SCET$_I$. The new intermediate theory is matched onto SCET$_{II}$ and the running of SCET$_I$ operators is performed to sum large perturbative logarithms. The values of soft-overlap function $\zeta_{\perp}$ for $K_1 (1270,\;1400), a_{1}$ and $b_{1}$ mesons are estimated from the light cone-sum-rules (LCSR), and later using it the corresponding branching fractions for $B \to \left(K_{1}(1270,\; 1400),\; a_{1},\; b_{1}\right)\gamma$ decays are calculated. We find that in case of $B \to K_{1}(1270,\; 1400)\gamma$ decays the results are in good agreement with their experimental measurements. Also the estimated values of the branching ratios of the $B \to (b_{1},\; a_1)\gamma$ decays are potentially large to be measured at the LHCb and future B-factories.


\end{titlepage}

\section{Introduction}\label{intro}
The rare decays involving $b \to s,\; d$ transitions are suppressed in the Standard Model (SM) due to Glashow-Iliopoulos-Maiani (GIM) mechanism. These flavor changing neutral current (FCNC) transitions make the rare decays important for the precision tests of the SM predictions and also to hunt for the physics beyond it, i.e., the new physics (NP). In this regard, the corresponding exclusive radiative process $B\rightarrow V\gamma$ (V is a vector meson) had been studied extensively both in SM \cite{Misiak:2015xwa,Ball:2006eu,Faustov:1992xv} and in various NP scenarios \cite{Oshimo:1992zd,Jung:2012vu}. Similarly, motivated by the above fact and the complementary confirmation the exclusive radiative decays $B\rightarrow A \gamma$, $A$ is any of the the axial-vector mesons  $K_1(1270),\; K_1 (1400),\; b_1(1235))$ and $a_1(1260)$, have also been studied in the context of large energy effective theory (LEET) in \cite{Lee:2004ju, Kwon:2004ri, JamilAslam:2005mc, Aslam:2006vh, JamilAslam:2006bw}. However, it is well known that among other effective field theories the power of SCET lies in the fact that it provides a platform to study the decays at different momentum scales. Along with this, it is a systematic field-theoretic approach that provides a freedom to study beyond leading order corrections, powers corrections (PC) as well as the running of the operators to sum the large logarithms. Therefore, we expect that the results obtained in this case will be more precise as compared to the other approaches, and also they will help us to get a clear understanding of the underlying physics. With this motivation, the present work is dedicated to the study of $B\rightarrow A \gamma$ decays in SCET which at a hadronic level corresponds to the heavy-to-light transitions.

In case of heavy-to-light decays, the standard SCET factorization is already developed [see for example \cite{Bauer:2001rd,Beneke:2003pa} for details and its proof]. The factorization relation for the matrix elements of the dimension-six operators $Q_i$ (c.f. Eq. (\ref{Q})) at leading power in the expansion $\Lambda/E$ (where $E$ is the final meson energy) and to all orders in $\alpha_s$ is
\begin{equation}
\langle A \gamma|Q_i|B\rangle = C_i ^{I}\zeta^{B\rightarrow A_{\perp}}(E)+\int_0 ^\infty \frac{d\omega}{\omega}\phi_B(\omega)\int_0 ^1 du \phi_{A_{\perp}}(u)C_i ^{II}(\omega,u)\label{theorem}
\end{equation}
where $C_i ^{I}$ and $C_i ^{II}$ are the Wilson coefficients (WCs) of the effective theory that we need to calculate using perturbation theory. The hadronic quantities $\phi_{A_\perp}$ and $\phi_{B}$ are the light-cone distribution amplitudes (LCDAs) for axial-vector and $B$ mesons, respectively. In the large recoil limit,  a single function $\zeta^{\perp} (E)$ - often called the soft overlap function is enough to describe the decay form factors of $B\rightarrow A_\perp \gamma$ up to factorizable corrections as shown in \cite{Charles:1998dr}. Along with this overlap function, we also require $\zeta^\parallel (E)$ in the semi-leptonic decays where $\perp$ and $\parallel$ correspond to the perpendicular and parallel polarizations of the final state meson, respectively.

 The general proof of writing the relation (\ref{theorem}) is based on the separation of soft and hard-scattering contributions (which are also known as the soft and hard modes) to the form factors, the convergence of convolution integral, and the absence of the contribution from non-valence Fock states at leading order.  The soft contributions in the overlap functions encapsulated in the matrix elements of SCET operators between initial and final state mesons that obey the spin-symmetry relations at the large-recoil. Such operators (in SCET) are not simple and lead to "non-factorizable" matrix elements that are sensitive to the end-point singularities and non-valence Fock states.  Therefore, the validity of Eq. (\ref{theorem}) lies in the fact that any non-factorizable part of the decay amplitude can be represented by $\zeta(E)$. 
 
 It is worthy to show that the diagrams in which the emitted photon from one of the quarks have the same structure as those of the diagrams of heavy-to-light form factors \cite{Beneke:2001at}. This proof gets complicated when the emitted photon is from the spectator quark of $B-$meson; because in such a scenario, some quarks and gluons will be collinear to the photon that can lead to the singularities in the momentum.  To cope with this issue, one needs to introduce new soft-collinear modes \cite{Becher:2003qh} in theory and has to show that one can decouple such modes in the low-energy theory at a leading- power. It is important to mention that a finite mass of $s-$quark would not spoil our factorization formula and contribute only in  $\phi_{A}^{\perp}$ and $\zeta_\perp^{B\rightarrow A_\perp}$. Here, we will ignore these contributions, which can be taken into account by following the method outlined in \cite{DeFazio:2007hw} for the explicit calculation of $\zeta_\perp ^{B\rightarrow(K_1, a_1, b_1)}$ and $\phi_{(K_1 ,a_1 ,b_1)}^\perp$.

In SCET, we deal with different momentum modes of quark and gluon fields, and the definition of physical scales becomes important for matching the effective theory. Therefore in this formalism, one can define the momentum modes according to Eq. (\ref{mom}) in soft, soft-collinear, collinear, hard, and hard-collinear modes.  Some of these modes are integrated-out while the others are required to preserve Eq. (\ref{theorem}). That is a two-step procedure accommodated by the expansion parameter $\lambda \sim \Lambda/m_b$. At a first step, we need to match the QCD to some intermediate effective theory, i.e., $\text{SCET}_I$ at hard scale $\mu_h \sim m_b$ to ensure that the hard modes $(1,1,1)m_b$ are integrated out. The next step is to match SCET$_I$ to a final theory called SCET$_{II}$ at an intermediate scale $\mu_i \sim \sqrt{m_b \Lambda}$. This ensures that the hard-collinear modes $(\lambda,1,\lambda^{1/2})m_b$ are also integrated out and we are left with the final theory of soft $(\lambda,\lambda,\lambda)m_b$, collinear $(\lambda^2 ,1,\lambda)m_b$ and soft-collinear $(\lambda^2, \lambda, \lambda^{3/2})m_b$ modes. The final thing that needed to be proven is whether the soft-collinear modes spoil the validity of Eq. (\ref{theorem}) or not. We have discussed this issue at length in Sec. \ref{sectioniii}.

In the two-step procedure mentioned above, it is pertinent to discuss the operators required at a given order in power counting. In this work, we are concerned with contributions to an-order of $\mathcal{O}(\alpha_s)$. It requires three types of operators satisfying Eq. (\ref{theorem}): the A-type operators that are quark-antiquark two-particle currents along with a photon emission from the quark-antiquark. The B-type operators are three particle quark-antiquark-gluon currents with photon emission from quark-antiquark, and the C-type operators are four fermion operators with a photon emitting from the spectator quark-antiquark. In our forthcoming study, we will discuss these three types of operators briefly. 

To bridge the matching between SCET$_I$ and SCET$_{II}$, the notion of hard $(\mu_h)$ and intermediate ($\mu_i$) scales are very useful. It will make it easy to sum the large perturbative logarithms in going from hard $\mu_h\sim m_b$ to an intermediate-scale $\mu_i\sim \sqrt{m_b \Lambda}$ by Renormalization Group (RG) analysis of SCET$_I$. Similarly, the RG-analysis of SCET$_{II}$ can be done in going from $\mu_i$ to a hadronic scale $\mu\sim \Lambda$ at which we wish to study the mesons. It will also require the running of meson distribution amplitude that in itself is a complicated procedure -  but the intermediate scale, in this case, is already close to final state masses, and hence it is avoided.

Regarding the meson in the final state, in the study presented here, we are interested in the axial-vector mesons $(K_{1B},\; b_1)$ and $(K_{1A},\; a_1)$ which are $^1P_1$ states with $J^{PC} = 1^{+-} $ and $1 ^3P_1$ states with $J^{PC}=1^{++}$, respectively. We also know that the physical axial-vector meson $(K_1(1270),\; K_1(1400))$ states arise as a result of mixing of $1 ^1P_1(K_{1A})$ and $1 ^3P_1(K_{1B})$ states. Consequently, these axial-vector mesons are relatively massive (around $1.3$ GeV) as compared to the vector meson states $(K^*,\omega,\rho)$. However, they are not heavy enough to rule out the heavy-quark-large-energy expansion that we use for the heavy-to-light decays. Recently, the branching fractions of $B \to K_1(1270,1400)\gamma$ decays are reported to be \cite{PDG:2020zy}:
\begin{eqnarray}
\mathcal{B}(B^+\rightarrow K_1 ^+(1270)\gamma)=(4.4^{+0.7}_{-0.6})\times 10^{-5},\notag\\
\mathcal{B}(B^+\rightarrow K_1 ^+(1400)\gamma)=(10^{+5}_{-4})\times 10^{-6}.\label{belle-value}
\end{eqnarray}
In the case of the final state axial-vector mesons  $(b_1,\;a_1)$, the experimental values of the branching ratios are still unknown. However, by looking at the CKM matrix elements involved in these decay, one can expect their branching ratios to be two orders of magnitude smaller than $B \to K_{1}(1270)\gamma$ decay, which is also the case for their corresponding vector meson states $(\omega)$ and $(\rho)$  \cite{Aubert:2006ag, Abe:2005rj}
\begin{eqnarray}
\mathcal{B}(B\rightarrow \omega\gamma)&=&( 5.4\pm 0.23(\text{stat.}) \pm 0.21(\text{syst.}))\times 10^{-7},\nonumber\\
\mathcal{B}(B^+\rightarrow \rho ^+\gamma)&=&( 6.8\pm 0.36(\text{stat.}) \pm 0.31(\text{syst.}))\times 10^{-7}.\label{bration-Vector}
\end{eqnarray}

The work performed here is organized as follows: In Sec. \ref{sectionii} we introduce the SCET$_I$ operators for (A, B, C)-type currents that are relevant to the radiative decays under consideration. In the same section, we work out the matching calculation from QCD to the SCET$_I$. In Sec. \ref{sectioniii}, the SCET$_{II}$ operators are given as four-quark operators to match with the SCET$_I$ operators. The relevant WCs, known as the jet-functions, are presented in the same section. We calculate the matrix elements for $B\rightarrow A\gamma$ in Sec. \ref{mat}, and for their completeness, we add the contributions of large perturbative logarithms. Sec1. \ref{res} is devoted to this discussion. Section \ref{branching} presents the results for the branching ratios along-with the calculations of soft overlap function using light-cone sum rules. Finally,  Sec. \ref{summary} provides a summary of our main results. The appendix contains our discussion of the matching contributions of the quark loop operator $Q_1$ and chromomagnetic operator $Q_8$, given in Eq. (\ref{Q}).

\section{SCET$_I$ Operators and QCD$\rightarrow$ SCET$_I$ matching}\label{sectionii}
The effective Hamiltonian corresponding to the FCNC $b\rightarrow (s,d)$ transition can be written as \cite{Buchalla:1995vs}
\begin{equation}
H_{eff}=-\frac{G_F}{\sqrt{2}}V_{tp}^{*} V_{tb}\sum_{i=1}^{8}C_i(\mu)Q_i(\mu),\label{Heff}
\end{equation}
with $p=s$ for $K_1$ and $d$ for $(b_1,\;a_1)$ mesons. Here, $C_i(\mu)$ and  $Q_i(\mu)$  with $i=1,...,8$ are the WCs and their corresponding operators, respectively. The operators $Q_i(\mu)$ are expressed as
\begin{eqnarray}
Q_1&=&\bar{s}\gamma^{\mu}(1-\gamma_5) c\quad \bar{c}\gamma_{\mu}(1-\gamma_5) b,\nonumber\\
Q_2&=&\bar{s}^{\alpha}\gamma^{\mu}(1-\gamma_5) c^{\beta}\quad\bar{c}^{\beta}\gamma_{\mu}(1-\gamma_5) b^{\alpha},\nonumber\\
Q_3&=&\bar{s}\gamma^{\mu}(1-\gamma_5) b\sum_q (\bar{q}\gamma_{\mu}(1-\gamma_5) q),\nonumber\\
Q_4&=&\bar{s}^{\alpha}\gamma^{\mu}(1-\gamma_5) b^{\beta}\sum_q(\bar{q}^{\beta}\gamma_{\mu}(1-\gamma_5) q^{\alpha}),\nonumber\\
Q_5&=&\bar{s}\gamma^{\mu}(1-\gamma_5) b\sum_q (\bar{q}\gamma_{\mu}(1+\gamma_5) q),\nonumber\\
Q_6&=&\bar{s}^{\alpha}\gamma^{\mu}(1-\gamma_5) b^{\beta}\sum_q(\bar{q}^{\beta}\gamma_{\mu}(1-\gamma_5) q^{\alpha}),\nonumber\\
Q_7&=&-\frac{e}{8\pi^2}m_b\bar{s}\sigma^{\mu\nu}(1+\gamma_5)b F_{\mu\nu},\nonumber\\\
Q_8&=&-\frac{g}{8\pi^2}m_b\bar{s}\sigma^{\mu\nu}(1+\gamma_5)T^a b G_{\mu\nu}^a ,\label{Q}
\end{eqnarray}
where $\alpha$ and $\beta$ represent the color indices and $T^a$ is a $SU(3)$ color matrix. The WCs in Eq. (\ref{Heff}) are known at next-to-next leading logarithm (NNLL) \cite{Bobeth:1999mk, Czakon:2006ss}. Though the matrix elements for the QCD penguin operators $Q_3 - Q_6$ arise at $\mathcal{O}(\alpha_s)$, which is compatible with our study - their relevant coefficients are too small to contribute and hence we can safely ignore them in our forthcoming analysis. The operator $Q_2$ contributes at $\mathcal{O}(\alpha_s ^2)$, so its contribution omitted too. Therefore, among the eight operators defined in Eq. (\ref{Q}), the phenomenologically relevant operators for the radiative decay $B\rightarrow A\gamma$ at $\mathcal{O}(\alpha_s)$ are $Q_1 ,Q_7$ and $Q_8$.

Before we proceed to the formal calculations, it is useful to adopt the general way of writing the momenta of quark or gluon fields in terms of light-like vectors, i.e., 
\begin{equation}
p^\mu = n\cdot p \frac{\bar{n}^\mu}{2}+\bar{n}\cdot p \frac{n^\mu}{2}+p_\perp ^\mu = p_+^\mu +p_- ^\mu +p_\perp ^\mu , \label{mom}
\end{equation}
Defining $P^\mu=m_Bv^\mu$, so that the four-vector $v^\mu$ is in the direction of motion of $B-$meson. As the axial-vector meson(s) is significantly lighter compared to $B-$meson, therefore, the light-like vector $n^\mu$ is in its direction of motion with the properties $n^2 = \bar{n}^2 = 0$ and $n\cdot\bar{n}=2$ along with $v_\perp =0$. The different scaling of the set $(p_+ , p_-, p_\perp)$ depicts different momentum modes which were already explained in the Sec. \ref{intro}.

In SCET, we can define the moment mode for each quark and gluon fields. In this regard, the SCET$_I$ is a theory that contains soft and hard-collinear modes. The hard-collinear quark and gluon fields along with heavy-quark and soft fields scale as
\begin{eqnarray}
\xi_{hc} , \xi_{\bar{hc}} \sim\lambda^{1/2},\quad A_{hc}^\mu \sim (\lambda , 1, \lambda^{1/2}),\quad h\sim \lambda^{3/2},\quad A_s\sim(\lambda,\lambda,\lambda).
\end{eqnarray}
To obtain the gauge-invariant operators for the fields defined in Eq. (\ref{Q}), we require light-like Wilson lines which connect them at different points. Therefore, in SCET$_I$, the hard-collinear fields should also be invariant under hard-collinear gauge transformations, which requires to introduce of a hard-collinear Wilson line
\begin{equation}
\mathcal{W}_{hc}(x)=\text{Pexp}\left(ig\int_{-\infty} ^0 ds \bar{n}\cdot A_{hc}(x+s\bar{n})\right), \label{Wiline}
\end{equation}
where the gauge-invariant hard-collinear fields are defined as
\begin{eqnarray}
\chi_{hc}(x)&=&W^\dagger_{hc}(x)\xi_{hc}(x),\nonumber\\
\mathcal{A}_{hc} ^\mu (x)&=&W^\dagger _{hc}[iD_{hc}^\mu (x)W_{hc}(x)]+\frac{\bar{n}^\mu}{2}[W^\dagger _{hc}(x)gn\cdot A_s(x_-)W_{hc}(x)-gn\cdot A_s(x_-)].
\end{eqnarray}
In above equations, it can be noticed that the position variable, $x$, is defined in a similar way as the light-like momentum, i.e., $x\equiv x_+ +x_- +x_\perp$ and $s$ is the light-ray variable in Eq. (\ref{Wiline}).
To construct operators at leading power using the above fields, it is pertinent to show that they are invariant under first reparmeterization conditions, i.e., $n^\mu \rightarrow (1+\alpha)n^\mu$ and $\bar{n}^\mu \rightarrow (1-\alpha)\bar{n}^\mu$. The tree level coefficients for A-type operators of the form $\bar{\chi}_{hc}\Gamma h$  with $\Gamma= I$ for scalar, $\Gamma_{V_i} = (\gamma ^\mu, v^\mu , n^\mu )$ for  vector currents, $\Gamma_{T_j} = (\gamma^{[\mu}\gamma^{\nu]},\gamma^{[\mu}v^{\nu]},\gamma^{[\mu}n^{\nu]},v^{[\mu}n^{\nu]})$ for tensor currents \cite{Hill:2004if,Beneke:2004rc}.
The corresponding tree level coefficients are
\begin{eqnarray}
C_S ^A &=&1,\quad C_{V1} ^A=1,\quad C_{V2,3} ^A=0,\quad C_{T1}^A=1,\quad C_{T2,3,4}^A =0. \label{treelevA}
\end{eqnarray}
The coefficients of pseudo-scalar A- type operators are similar to the scalar ones while the coefficients of axial-vector A-type operators are related to the vector ones as  $C_{V1,V2}^A=C_{A1,A2}^A$ and $C_{V3}^A=-C_{A3}^A$.  As the operators in the above basis mix up B-type operators of the form $\bar{\chi}_{hc}\Gamma\slashed{\mathcal{A}}_{hc}h$, the matching results are obtained by using a new basis (for details see ref. \cite{Hill:2004if}). 
The B-type operators for vectors $\Gamma_{i}^\mu=(\gamma_\perp^\mu,v^\mu ,n^\mu)$ and tensors $\Gamma_j ^{\mu\nu}=(\gamma^{[\mu}_\perp \gamma^{\nu]}_\perp,\gamma^{[\mu}_\perp v^{\nu]},\gamma^{[\mu}_\perp n^{\nu]},v^{[\mu}n^{\nu]})$ in new basis along with their tree level coefficients read as follows
\begin{eqnarray}
J_{S}^{B'}&=&\bar{\chi}_{hc}(s\bar{n})\slashed{\mathcal{A}}_\perp(r\bar{n})h(0),\qquad\qquad\qquad\ \ \ \  C_S ^{B'}=-1,\nonumber\\
J_{Vi}^{B'\mu}&=&\bar{\chi}_{hc}(s\bar{n})\slashed{\mathcal{A}}_\perp(r\bar{n})\Gamma_i^\mu h(0),\qquad\qquad\qquad C_{V1}^{B'}=1,C_{V2}^{B'}=-2,C_{V3}^B=1-z,\nonumber\\
J_{V4}^{B'\mu}&=&\bar{\chi}_{hc}(s\bar{n})\gamma_\perp ^\mu\slashed{\mathcal{A}}_\perp(r\bar{n}) h(0),\qquad\qquad\qquad C_{V4}^B=0,\nonumber\\
J_{Tj}^{B'\mu}&=&\bar{\chi}_{hc}(s\bar{n})\slashed{\mathcal{A}}_\perp(r\bar{n})\Gamma_j^{\mu\nu} h(0),\qquad\qquad\qquad\  C_{T1}^{B'}=-1, C_{T2}^{B'}=-4,C_{T3}^{B'}=2,C_{T4}^{B'}=2,\nonumber\\
J_{T5}^{B'\mu}&=&\bar{\chi}_{hc}(s\bar{n})\mathcal{A}_{\perp\alpha}(r\bar{n})\gamma_\perp^{[\alpha}\gamma^\mu_\perp\gamma^{\nu]}_\perp h(0),\qquad\ \ \ \ \ \  C_{T5}^{B'}=0,\nonumber\\
J_{T6}^{B'\mu}&=&\bar{\chi}_{hc}(s\bar{n})v^{[\mu}\gamma_\perp^{\nu]}\slashed{\mathcal{A}}_{\perp}(r\bar{n}) h(0),\qquad\qquad\qquad C_{T6}^{B'}=0,\nonumber\\
J_{T7}^{B'\mu}&=&\bar{\chi}_{hc}(s\bar{n})n^{[\mu}\gamma_\perp^{\nu]}\slashed{\mathcal{A}}_{\perp}(r\bar{n}) h(0),\qquad\qquad\qquad C_{T7}^{B'}=2z , \label{treelev}
\end{eqnarray}
where $z=2E/m_b$ with $E$ being the energy of the final state quark and $\gamma_\perp ^\mu = \gamma^\mu-n^\mu\slashed{\bar{n}}/2-\bar{n}^\mu\slashed{n}/2$. For B-type operators, the coefficients of scalar and psuedoscalar currents have opposite sign whereas the coefficients of axial-vector currents are related with the vector one as $C^{B'}_{V1,V4}=-C^{B'}_{A1,A4}$ and $C^{B'}_{V2,V3}=C^{B'}_{A2,A3}$.

To construct the operators in the SCET$_I$ for $B\rightarrow A\gamma$ decay; it is required to have one $n$-hard-collinear field (axial-vector meson in our case), one $\bar{n}$-hard-collinear field ($\gamma$ in our case), and a heavy quark field. So at leading power in $\lambda$, the relevant operator and WC for A-type operator is
\begin{eqnarray}
J ^A (s,a) &=& \bar{\chi}_{hc}(s\bar{n})(1+\gamma_5)\slashed{\mathcal{A}}_{\bar{hc}\perp}^{(em)} h(0),\nonumber\\ \label{Aoper}
C^A(E,E_\gamma)&=&\int ds\int da e^{is\bar{n}\cdot P}e^{ian\cdot P_\gamma}\tilde{C}^A(s,a),
\end{eqnarray}
 where $E =  \bar{n}\cdot P/2$ and $E_\gamma\equiv n\cdot P_\gamma /2$. Also, $\bar{n}\cdot P$ is the large component of the total outgoing $n$-hard-collinear momentum and $n\cdot P_\gamma$ is the outgoing photon momentum. Similarly, the relevant SCET$_I$ operators for B-type currents (three particle current) and their corresponding WCs are
\begin{eqnarray}
J^{B}_1(s,r,a)&=&\bar{\chi}_{hc}(s\bar{n})(1+\gamma_5)\slashed{\mathcal{A}}_{\bar{hc}\perp}^{(em)}(an)\slashed{\mathcal{A}}_{hc\perp}(r\bar{n}) h(0),\label{Boper1}\\
J^{B}_2(s,r,a)&=&\bar{\chi}_{hc}(s\bar{n})(1+\gamma_5)\slashed{\mathcal{A}}_{hc\perp}(r\bar{n})\slashed{\mathcal{A}}_{\bar{hc}\perp}^{(em)}(an) h(0),\label{Boper2}\\
C^B _j(E,E_\gamma,u)&=&\int ds\int dr\int da e^{i(us+\bar{u}r)\bar{n}\cdot P}e^{ian\cdot P_\gamma}\tilde{C}_j ^B(s,r,a).
\end{eqnarray}
The index $j=1,\;2$ in B-type WCs refers to the two types of operators written in Eqs. (\ref{Boper1}, \ref{Boper2}). The variables $u$ and $\bar{u}$ denote the fraction of the large component of the $n$-hard-collinear momentum carried by the quark and the gluon fields, respectively. The relevant SCET$_I$ operator for the photon emission from the spectator quark (C-type matching) is a four-quark operator. Along with the corresponding WC, it becomes
\begin{eqnarray}
J^C(s,r,a)&=&\bar{\chi}_{hc}(s\bar{n})(1+\gamma_5)\frac{\slashed{\bar{n}}}{2}\chi_{hc}(r\bar{n})\bar{\chi}_{\bar{hc}}(an)(1+\gamma_5)\frac{\slashed{n}}{2} h(0),\nonumber\\\label{Coper1}
C_k ^C(u)&=&\int ds\int dr\int da e^{i(us+\bar{u}r)\bar{n}\cdot P}e^{ian\cdot P_\gamma}\tilde{C}_k ^C(s,r,a).
\end{eqnarray}

\begin{figure}
\begin{subfigure}{.3\textwidth}

  \includegraphics[width=.9\linewidth]{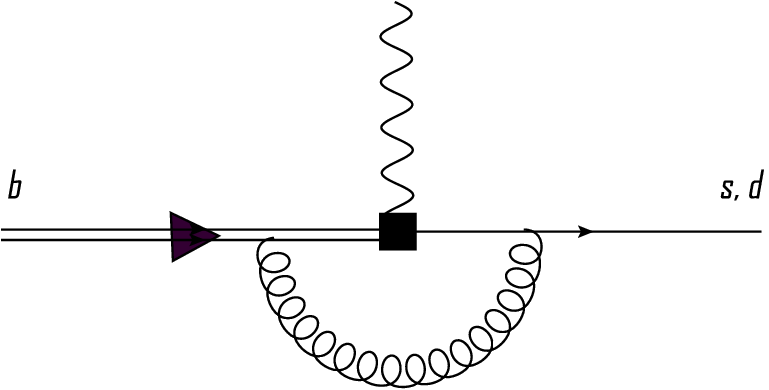}
  \caption{}
  \label{q7a}
\end{subfigure}%
\begin{subfigure}{.3\textwidth}

  \includegraphics[width=.9\linewidth]{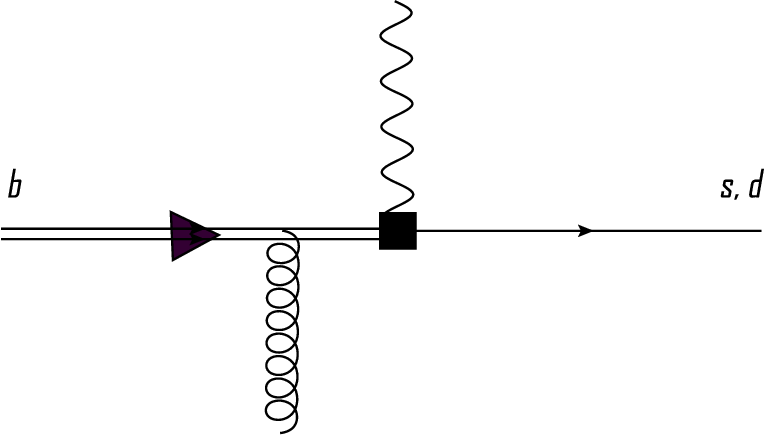}
  \caption{}
  \label{q7b1}
\end{subfigure}
\begin{subfigure}{.3\textwidth}

  \includegraphics[width=.9\linewidth]{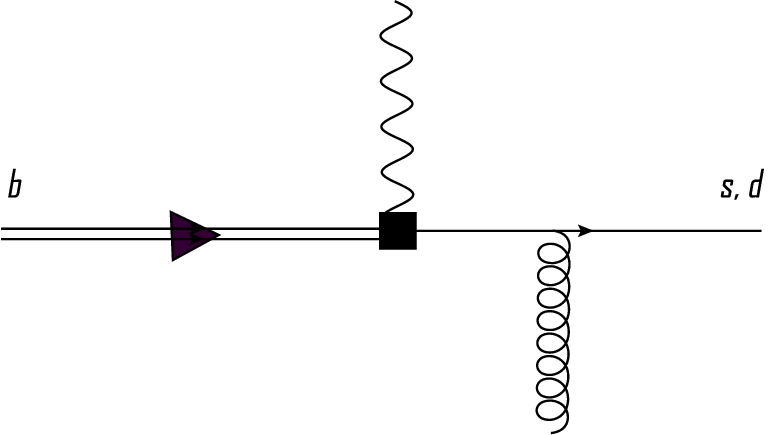}
  \caption{}
  \label{q7b2}
\end{subfigure}
\caption{The diagrams (a), (b) and (c) for matching $Q_7$ onto $J^A$ , $J_1^B$  and $J_2^B$, respectively.}
\label{fig:fig1}
\end{figure}

While matching QCD to SCET$_I$ at order $\alpha_s$, one needs to calculate the WCs at one-loop for the A-type operators while at tree level for the B-type currents. 
Let us denote $\Delta_i C_j^{(A,B,C)}$ for the purpose of matching the results of weak effective operators $Q_i$ to the SCET$_I$ operators $J_j ^{A,B,C}$. Thus, we can write the total matching coefficients as
\begin{eqnarray}
\mathcal{C} ^{(A,C)}&=&\frac{G_F}{\sqrt{2}}\sum_{p=u,c}V_{ps}^{*}V_{pb}\left[C_1(\mu_{_{QCD}})\Delta_1 ^p C^{(A,C)}(\mu_{_{QCD}},\mu) + C_7(\mu_{_{QCD}})\Delta_7  C^{(A,C)}(\mu_{_{QCD}},\mu)+C_8(\mu_{_{QCD}})\Delta_8  C^{(A,C)}(\mu_{_{QCD}},\mu) \right],\nonumber\\
\mathcal{C} ^{B}&=&\frac{G_F}{\sqrt{2}}\sum_{p=u,c}V_{ps}^{*}V_{pb}\left[C_1(\mu_{_{QCD}})\Delta_1 ^p C^{B}_{1,2}(\mu_{_{QCD}},\mu) +C_7(\mu_{_{QCD}})\Delta_7  C^{B}_{1,2}(\mu_{_{QCD}},\mu)+C_8(\mu_{_{QCD}})\Delta_8  C^{B}_{1,2}(\mu_{_{QCD}},\mu)  \right],
\end{eqnarray}
where $\mu_{_{QCD}}=\mu_h\sim m_b$ is the hard scale at which QCD is matched on to SCET$_I$ and $\mu\sim\sqrt{m_b\Lambda}$ is an intermediate scale. The tree-level matching of $Q_7$ is trivial as can be seen from Eq. (\ref{treelev}). The matching coefficient at order $\mathcal{O}(\alpha_s)$ can be obtained via matching the QCD diagram \ref{q7a} at leading power to SCET$_I$ current $J^A$ as given in \cite{Beneke:2004rc,Bauer:2000yr}
\begin{figure}
\begin{subfigure}{.3\textwidth}

  \includegraphics[width=.9\linewidth]{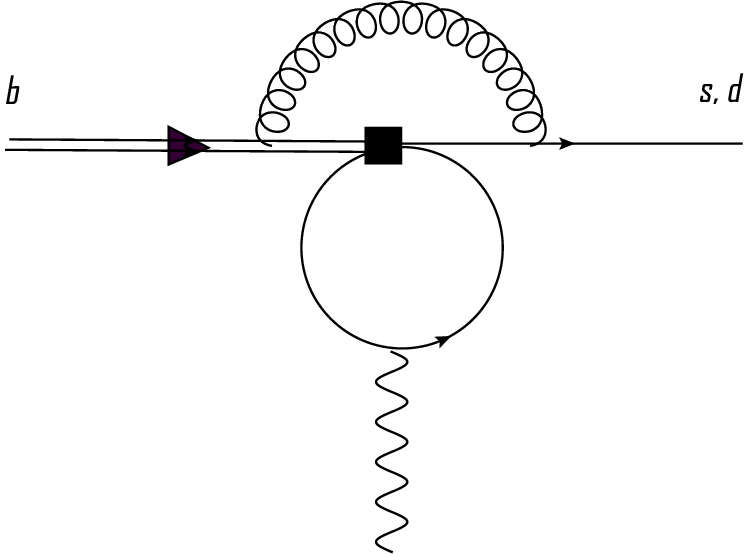}
  \caption{}
  \label{q1a}
\end{subfigure}%
\begin{subfigure}{.3\textwidth}

  \includegraphics[width=.9\linewidth]{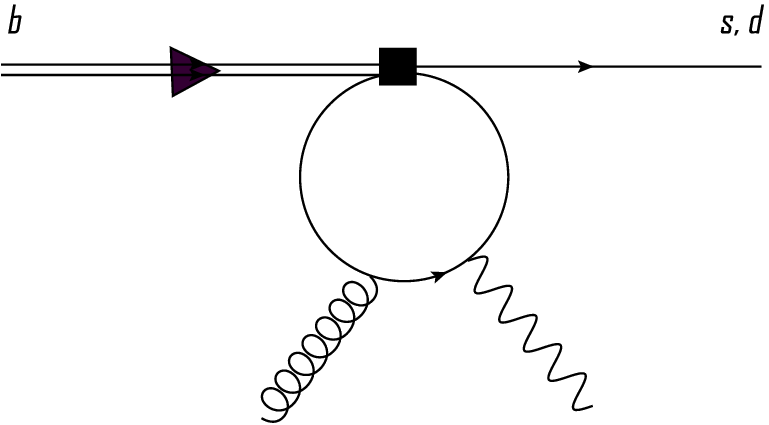}
  \caption{}
  \label{q1b1}
\end{subfigure}
\begin{subfigure}{.3\textwidth}

  \includegraphics[width=.9\linewidth]{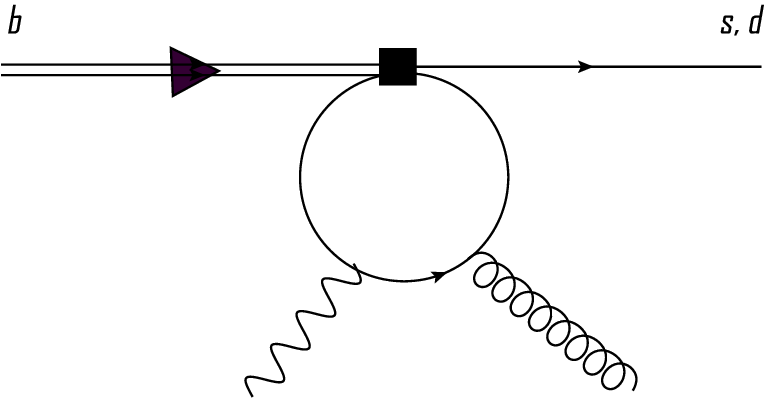}
  \caption{}
  \label{q1b2}
\end{subfigure}
\caption{The diagrams (a), (b) and (c) for matching $Q_1$ onto $J^A$ , $J_1^B$  and $J_2^B$, respectively.}
\label{fig:fig2}
\end{figure}
\begin{figure}
\begin{subfigure}{.4\textwidth}

  \includegraphics[width=0.95\linewidth]{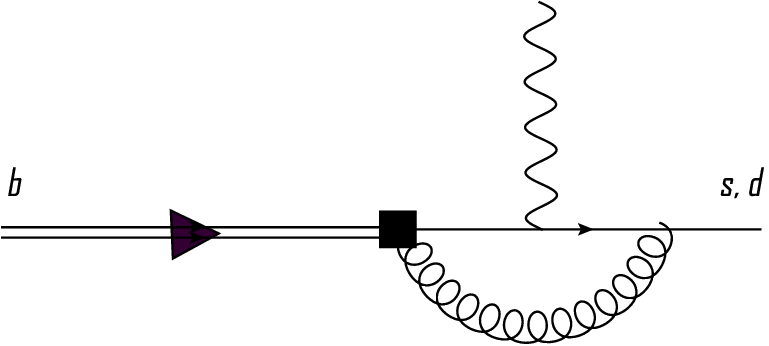}
  \caption{}
  \label{q8a}
\end{subfigure}%
\begin{subfigure}{.4\textwidth}

  \includegraphics[width=0.95\linewidth]{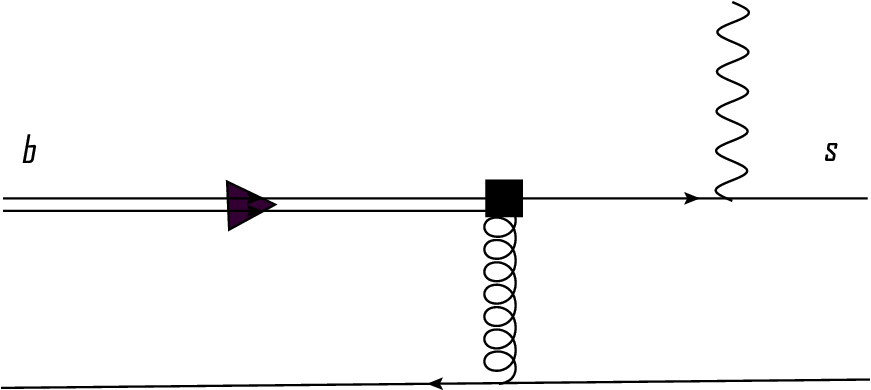}
  \caption{}
  \label{q8b}
\end{subfigure}

\caption{The diagrams (a), (b) for matching $Q_8$ onto SCET$_I$ operators.}
\label{fig:fig3}
\end{figure}

\begin{eqnarray}
\Delta_7 C^A &=&\frac{e\bar{m}_b E_\gamma}{2\pi^2}\left\lbrace 1 +\frac{\alpha_s(m_b)C_F}{4\pi}\left[-2 \text{ln}^2\frac{\mu }{2E} -5\text{ln}\frac{\mu}{2E}-2\text{ln}\frac{\mu_{_{QCD}}}{2E}- 2\text{Li}_2(1-\frac{2E}{m_b})-6-\frac{\pi^2}{12}\right]+\mathcal{O}(\alpha_s ^2)\right\rbrace,
\end{eqnarray}
where $m_b$ is the pole mass of $b-$quark and $\bar{m}_b$ is the running mass at $\mathcal{O}(\alpha_s)$ in $\overline{MS}$ scheme at the scale $\mu_{QCD}$, i.e.,
\begin{eqnarray}
\overline{m}_b(\mu_{QCD})&=& m_b\left[1+\frac{\alpha_s C_F}{4\pi}\left(3\text{ln}\frac{m_b^2}{\mu^2_{QCD}}-4\right)\right].\nonumber
\end{eqnarray}

The B-type operators defined in equations (\ref{Boper1}) and (\ref{Boper2}) are power suppressed as compared to A-type - but they will contribute at leading power when matched to the SECT$_{II}$ operators. For the coefficients of operator $Q_7$ one could match the QCD diagrams \ref{q7b1} and \ref{q7b2} to currents $J_1 ^B$ and $J_2 ^B$, respectively. Making use of the tensor coefficients for B-type currents in Eq. (\ref{treelev}), the matching gives
\begin{equation}
\Delta_7 C_1^B=\frac{e\bar{m}_b E_\gamma}{4\pi^2 m_b}\qquad, \qquad\Delta_7 C_2^B=0 +\mathcal{O}(\alpha_s).
\end{equation}

For the operator $Q_1$ with a $c-$quark loop (setting $m_u=0$), the QCD diagram that contributes at leading order for the A-type current matching is given in Fig. \ref{q1a}. It is a vertex diagram from which hard modes have to be integrated out along with the contribution of the fermion loop when matching to SCET$_I$. Figs. \ref{q1b1} and \ref{q1b2} are bremsstrahlung diagrams with an on-shell photon. The corresponding coefficients in this case are
\begin{eqnarray}
\Delta_1^c C^A &=&\frac{\alpha_s C_F}{4\pi}G_i(x_c)\Delta_7 C^A, \nonumber\\
\Delta_1^c C^B _1 (u) &=&-\Delta_1 ^q C_2 ^B(u)=\frac{2e}{3}f\left(\frac{\bar{m}_c ^2}{4\bar{u}EE_\gamma}\right) \Delta_7 C_1 ^B,
\end{eqnarray}
where $x_c =\bar{m}_c ^2/m_b^2$ and the functions $G_1(x_c)$ and $f(\frac{m_c^2}{4EE_\gamma\bar{u}})$ are given in the Appendix.

Ignoring the penguin operators $Q_3 - Q_6$, one is left with the gluon chromomagnetic operator $Q_8$. The diagrams shown in Figs. Figs. \ref{q8a} and \ref{q8b} are required to match the SCET$_I$ A-type and B-type operators, respectively. Contrary to this, the diagrams drawn in Fig. \ref{fig:fig3}, where the photon is emitted from the $b-$quark are suppressed therefore $\Delta_8 C_2 ^B (u)\simeq0$. The coefficients for the A- and B- type currents matching are \cite{Greub:1996tg}
\begin{eqnarray}
\Delta_8 C^A\ \ \ =\frac{\alpha_s C_F}{4\pi}G_8 \Delta_7 C^A\qquad;\qquad
\Delta_8 C_1 ^B (u)= \frac{\bar{m}_b}{4\pi^2}\frac{e}{3}\frac{\bar{u}}{u}.
\end{eqnarray}

\begin{figure}
\begin{subfigure}{.3\textwidth}

  \includegraphics[width=.7\linewidth]{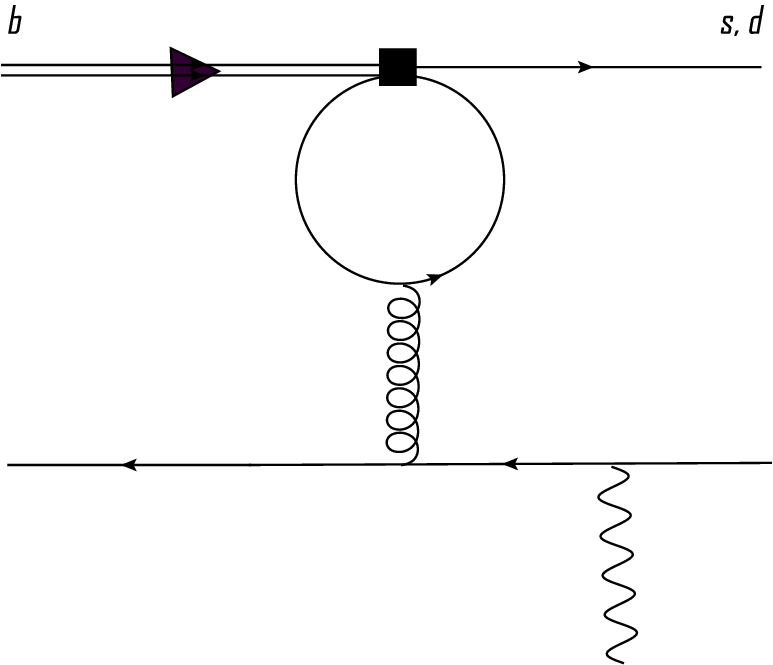}
  \caption{}
  \label{spec1}
\end{subfigure}%
\begin{subfigure}{.3\textwidth}
  \includegraphics[width=.7\linewidth]{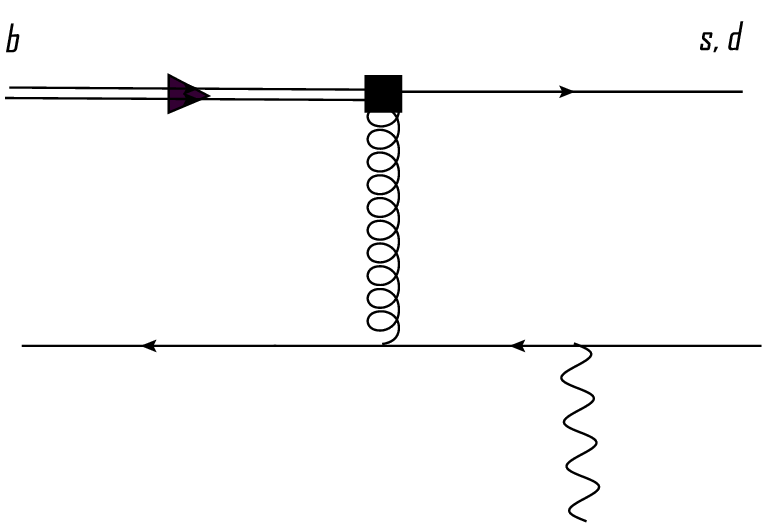}
  \caption{}
  \label{spec2}
\end{subfigure}

\begin{subfigure}{.3\textwidth}
  \includegraphics[width=.7\linewidth]{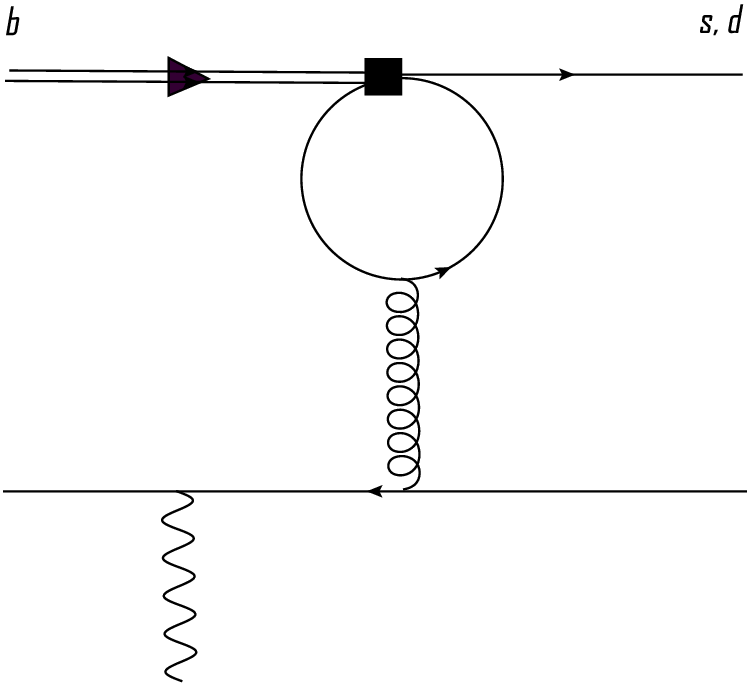}
  \caption{}
  \label{spec3}
\end{subfigure}
\begin{subfigure}{.3\textwidth}
  \includegraphics[width=.7\linewidth]{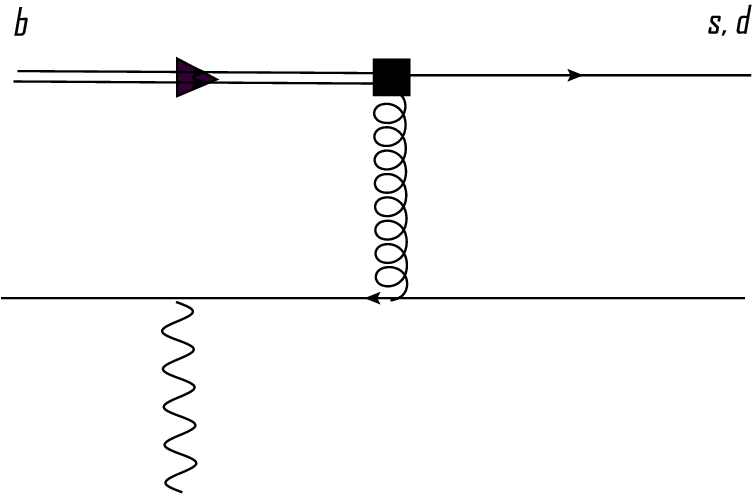}
  \caption{}
  \label{spec4}
\end{subfigure}
\caption{Diagrams with photon emission from the spectator anti-quark matched on to $J^C$.}
\label{fig:fig5}
\end{figure}

Now the photon emission from the spectator quark may contribute to the leading power. In Figs. \ref{spec1} and \ref{spec2}, the photon emission from anti-quark contributes at sub-leading power and hence is suppressed \cite{Chay:2003kb,DescotesGenon:2004hd}. In Figs. \ref{spec3} and \ref{spec4}, the photon-emission from spectator quark contributes at leading power when matched onto SCET$_{II}$ - but it vanishes when we calculate the matrix elements for the transversely polarized axial-vector mesons using Eq. (\ref{klcda}).


\section{SCET$_{II}$ Operators and SCET$_I$ $\rightarrow$ SCET$_{II}$ matching}\label{sectioniii}

We know that the SCET$_{II}$ is a low energy theory of soft, collinear, and soft-collinear modes. The next step in a two-step matching procedure is to go from an intermediate scale to the hadronic scale achieved by matching SCET$_I$ operators with SCET$_{II}$ operators. It requires the construction of SCET$_{II}$ operators at leading power scaling in a way we constructed them in the SCET$_I$. In these operators, the relevant WCs have known to be the Jet functions.  Furthermore, at a leading-power, the soft-collinear modes could be de-coupled via field redefinition in soft and collinear Lagrangians.
To obtain the gauge-invariant operators, again we need to define soft and collinear Wilson lines as we wrote them for hard-collinear fields in Eq. (\ref{Wiline})
\begin{eqnarray}
\mathcal{W}_{c}(x)&=\text{P exp}\left(ig\int_{-\infty} ^0 ds \bar{n}\cdot A_{c}(x+s\bar{n})\right),\nonumber\\
\mathcal{S}(x)&= \text{P exp}\left(ig\int_{-\infty} ^0 ds \bar{n}\cdot A_{s}(x+s\bar{n})\right).
\end{eqnarray}

It is useful to introduce the gauge-invariant quark and gluon fields which scales as follows
\begin{eqnarray}
\bar{\chi}_c &= \mathcal{W}_c ^{\dagger}\xi_c \sim \lambda ,\qquad Q_s = \mathcal{S}^\dagger q_s \sim \lambda ^{3/2}, \qquad \mathcal{H}= \mathcal{S}^\dagger h \sim \lambda^{3/2}, \nonumber\\
\mathcal{A}_c ^\mu &= \mathcal{W}_c ^{\dagger}(i D_c ^\mu \mathcal{W}_c)\sim (\lambda^2 ,0,\lambda), \qquad \mathcal{A}_s ^\mu = \mathcal{S} ^{\dagger}(i D_s ^\mu \mathcal{S})\sim (\lambda^2 ,0,\lambda).
\end{eqnarray}
We are not required to perform SCET$_I$ to SCET$_{II}$ matching for A-type currents as the matrix elements of $J^A$ already give the non-factorizable part, i.e., soft-overlap function.  Therefore we only need to define B-type operators for the matching. Recall, there were two B-type currents, so they match onto as
\begin{eqnarray}
O_1 ^B(s,t)&=&\bar{\chi}_c (s\bar{n})(1+\gamma_5)\slashed{\mathcal{A}}_{\bar{c}\perp}^{(em)}(0)\frac{\bar{\slashed{n}}}{2}\chi_c (0)\bar{\mathcal{Q}}_s (tn)(1-\gamma_5)\frac{\slashed{n}}{2}\mathcal{H}_s (0),\nonumber\\
O_2 ^B(s,t)&=&\bar{\chi}_c (s\bar{n})(1+\gamma_5)\frac{\bar{\slashed{n}}}{2}\chi_c (0)\bar{\mathcal{Q}}_s (tn)(1+\gamma_5)\frac{\slashed{n}}{2}\slashed{\mathcal{A}}_{\bar{c}\perp}^{(em)}(0)\mathcal{H}_s (0).\label{scet2op}
\end{eqnarray}

The photon field is in the $\bar{n}$ direction and is collinear. The first operator in Eq. (\ref{scet2op}) matches SCET$_I$ operator with photon emission from $b-$quark and it requires opposite chirality of the spectator quark field, $\bar{\mathcal{Q}}_s$. On the other hand, the second operator matches SCET$_I$ operator with photon emission from $s-$quark.

We can see that at leading power, the SCET$_{II}$ operators have a soft and a collinear part only. Therefore, the matrix elements would be found independently in terms of LCDA of $B-$ and final state axial-vector mesons. The decoupling of soft-collinear modes made our factorization successful. Moreover, if they did not decouple they would have appeared as end-point divergences in convolution integrals (\ref{theorem}), which is in contrast to the LEET as discussed in \cite{Sikandar:2019qyb}. To mention, the LEET does not correctly produce the infrared divergences; therefore, we left with end-point divergences arising from the convolution integral at a leading twist. These end-point divergences were then assumed to be absorbed in the soft-form factor in the LEET. Hence, it would suggest the difference between soft-form factors calculated in both theories.

 With this note, the corresponding WC in momentum space is given as
\begin{equation}
D_{1,2} ^{B}(\omega, u)\equiv \int ds \int dt e^{-i\omega n\cdot v t}e^{ius\bar{n}\cdot P}\tilde{D}_{1,2} ^B (s,t),
\end{equation}

\begin{figure}
  \includegraphics[width=100mm]{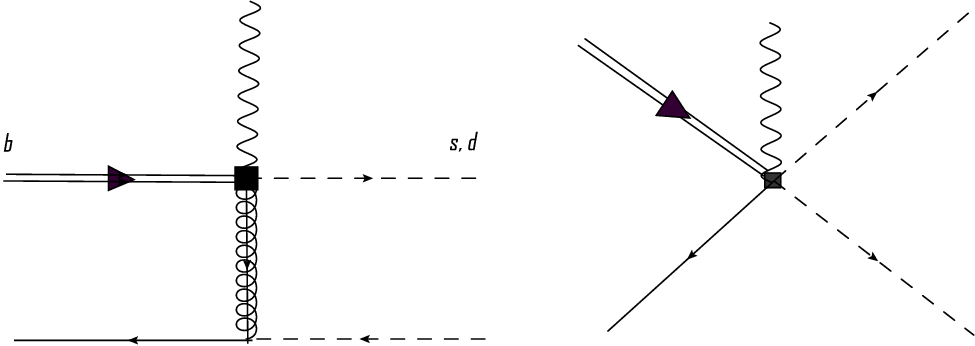}
\caption{SCET$_I$ diagram matching onto SCET$_{II}$ 4-quark operator. The dashed gluon is hard-collinear}.
\label{fig:fig6}
\end{figure}
which is actually a convolution of the SCET$_I$ WC, $C_i ^B$, with a jet function $\mathcal{J}_{\perp,\parallel}$, i.e.,
\begin{eqnarray}
D_{1,2} ^B(\omega, u, \mu_i)=\frac{1}{\omega}\int_0 ^1 dy \mathcal{J}_{\perp} \left(u,y,ln\frac{2E\omega}{\mu_i ^2 },\mu_i\right)C_{1,2} ^B(y,\mu_1),
\end{eqnarray}
where $\mu_i\sim \sqrt{2E\Lambda_{QCD}}$ is an intermediate scale. At tree level, the matching of $J_1 ^B$ onto $O_1 ^B$ is trivial (c.f. Fig. \ref{fig:fig6}) and the corresponding jet function is given as
\begin{eqnarray}
\mathcal{J}_\perp (u,v)&=&\mathcal{J}_\parallel (u,v)= -\frac{4\pi C_F \alpha_s}{N}\frac{1}{2E\bar{u}}\delta(u-v),
\end{eqnarray}
while the similar matching for $J_2 ^B$ is suppressed at a leading power. 

\section{Matrix Elements}\label{mat}
The quark and gluon fields defined so far are in different momentum regions. It will help us to conveniently match the effective theories at different scales and summing large logarithms. We will discuss it in detail in Sec. \ref{res}. As we have already mentioned that the $B-$meson is a bound state of two soft quarks, and the final state axial-vector meson is that of two collinear quarks; therefore, their four-momenta can be defined as
\begin{eqnarray}
p_B^\mu &=m_B v^\mu , \quad\quad p^\mu _A= E n^\mu +\frac{m_A ^2}{4E}\bar{n}^\mu\quad\quad p_{\gamma} ^\mu =( E-\frac{m^2 _A}{4E} )\bar{n}^\mu,
\end{eqnarray}
where $E$ is the off-shell energy with $p_A^2 =m_A ^2$ and $p_{\gamma} ^2 =0$. One can notice that the large component of the axial-vector meson is in the $n-$direction while the emitted photon in $\bar{n}-$direction. Similarly, the on-shell energy, the three-momentum of the axial-vector meson, and the energy of an on-shell photon are defined as
\begin{eqnarray}
E_F=\frac{m_B ^2 +m_{A}^2}{2m_B},\quad\quad |\bold{\Delta}| =\sqrt{E_F ^2 -m_{A}^2},\quad\quad  E_\gamma = \frac{m^2 _B -m^2_A}{2m_B},
\end{eqnarray}
where $E_F\sim E\sim\Delta= m_B/2$ is the energy of the final state meson at maximum recoil.  Let $\eta^*$ define the polarization vector of the axial-vector meson so that the transversality condition $\eta^*\cdot p_A =0$ gives
\begin{eqnarray}
\eta^{*}\cdot n =-\frac{m_{_{A}}^2}{4E\Delta}\eta^{*}\cdot \bar{n}.\label{identity}
\end{eqnarray}
The photon has polarization $\varepsilon^*$ and hence $\varepsilon^*\cdot p_\gamma=\varepsilon^*\cdot \bar{n} =0$. The $B\rightarrow A$ form factors can be parameterized in terms of soft-overlap function $\zeta_{A\perp}$ by using SCET$_I$ A-type operator as
\begin{eqnarray}
\langle A(p_A) |\bar{\chi}_{_{hc}}\Gamma h| B(v)\rangle &=-2E_F \zeta_{A}(E_F)\text{tr}[\bar{\mathcal{M}}_{A\perp}(n)\Gamma\mathcal{M}_B (v)],\label{matrixelements}
\end{eqnarray}
where the projection operators $\mathcal{M}_{A\perp}(n)$ and $\mathcal{M}_B (v)$ are
\begin{eqnarray}
\mathcal{M}_B (v)&=-\frac{1+\slashed{v}}{2}\gamma_5,\qquad \mathcal{\bar{M}}_{A \perp} (n)=-\slashed{\eta}_\perp ^*\gamma_5 \frac{\slashed{\bar{n}}\slashed{n}}{4}.
\label{proj}
\end{eqnarray}
Substituting these projectors from Eq. (\ref{proj}) in Eq. (\ref{matrixelements}) with $\Gamma = (1+\gamma_5)$, we get
\begin{eqnarray*}
\langle A(p_F,\eta^*)\gamma(\varepsilon^*) |J^A| B(v)\rangle =\mathcal{C}^A E_F \zeta_{A\perp}(E_F)\left[2 (\varepsilon^{*}\cdot \eta^*)-(n\cdot\eta^*)(\bar{n}\cdot\varepsilon^*)\right. \\
\left. -(n\cdot\varepsilon^*)(\bar{n}\cdot\eta^*)+i\epsilon^{\alpha\beta\mu\nu}n_\alpha\bar{n}_{\beta}\varepsilon^{*}_\mu\eta^*_\nu\right].
\end{eqnarray*}
As $\bar{n}\cdot\varepsilon^*=0$ and because both photon and axial-vector meson are left circularly polarized, $\varepsilon^* \cdot \eta^* =1$. Making use of Eq. (\ref{identity}) and $\epsilon^{0123} =-1$, we get $\epsilon^{\alpha\beta\mu\nu}n_\alpha\bar{n}_{\beta}\varepsilon^{*}_\mu\eta^*_\nu =-2i$. This leads to
\begin{eqnarray}
\langle A(p_F)\gamma |J^A| B(v)\rangle =4 \mathcal{C}^AE_F\left(1+\frac{m^2_{A}}{8E\Delta}\right)\zeta_{A\perp}(E_F).\label{Aresult}
\end{eqnarray}
The matrix elements of B-type currents $O_{1,2}^B$ of SCET$_{II}$, with independent soft and collinear parts can be written as a convolution of the LCDA's of the respective meson given here as 
\begin{eqnarray}
\langle 0| \bar{\mathcal{Q}}_s(tn)\frac{\slashed{n}}{2}\Gamma \mathcal{H}(0)|\bar{B}(v)\rangle &=&-\frac{i\sqrt{m_B}F(\mu)}{2}\text{Tr}\left(\frac{\slashed{n}}{2}\Gamma\frac{1+\slashed{v}}{2}\gamma_5 \right)\times\int_0 ^\infty d\omega e^{-i\omega t n\cdot v }\phi_B (\omega, \mu),\label{blcda}\\
\langle A_{\perp}(p_F,\eta^*)|\bar{\chi}_c (s\bar{n})\Gamma \frac{\bar{n}}{2}\chi_c (0)|0\rangle&=&-\frac{if_{A_{\perp}}(\mu)}{4}\bar{n}\cdot p_F \text{Tr}\left(\slashed{\eta}^*_\perp\gamma_5 \Gamma \frac{\slashed{\bar{n}}\slashed{n}}{4}\right)\times\int_0 ^\infty du e^{ius\bar{n}\cdot p_F }\phi_{A_{\perp}}(u,\mu),\label{klcda}
\end{eqnarray}
where $\phi_B$ and $\phi_{A\perp}$ are the distribution amplitudes for $B-$ and $A-$mesons, respectively, and $f_{A\perp}$ is the axial-vector meson decay constant which depends upon the scale of the theory. The scale-dependent quantity $F(\mu)$ is related to $B-$meson decay constant $f_B$ at NLO as
\begin{eqnarray*}
f_B\sqrt{m_B}&=F(\mu)\left(1+\frac{C_F \alpha_s (\mu)}{4\pi}\left(3\text{ln}\frac{m_b}{\mu}-2\right)\right).
\end{eqnarray*}
For $\Gamma = (1-\gamma_5)$ the $B-$meson LCDA  is $n\cdot v$ and for  $\Gamma =(1+\gamma_5)\gamma_\perp ^\mu$ it vanishes. Contrary to this, for $\Gamma = (1+\gamma_5)$ axial-vector meson LCDA vanishes, whereas for $\Gamma = (1+\gamma_5)\gamma_\perp ^\mu$ it is $(1+m_{A}^2/8E\Delta)$ and hence $O_1 ^B$ will effectively contributes. Collecting the results from Eqs. (\ref{Aresult}, \ref{blcda}) and Eq. (\ref{klcda}) to get the complete matrix elements at leading power and at an order $\mathcal{O}(\alpha_s)$, we have
\begin{eqnarray}
\langle A_{\perp}(p_F,\eta^*)\gamma(\varepsilon^*)|\mathcal{H}_W |B(v)\rangle &=& 4 \left(1+\frac{m^2 _A }{8E\Delta}\right)\left[ \mathcal{C}^A E_F\zeta_{A\perp} (E_F)
-\frac{m_B ^{3/2} F(\mu)}{2}\int_0 ^\infty \frac{d\omega}{\omega}\phi_B (\omega, \mu)\right.\nonumber\\
&\times&\left. \int_0 ^1 du  f_{A\perp}\phi_{A_{\perp}}(u,v)\int_0^1dv \mathcal{J}_\perp \left(u,v,\text{ln}\frac{m_B \omega}{\mu^2}\right)\mathcal{C}_1 ^B(v,\mu)\right].\label{matrixel}
\end{eqnarray}

\section{Resummation}\label{res}

The matching coefficients calculated in the previous section are reliable and can be trusted up-to a hard-scale $\mu_h\sim m_b$. However, one needs to re-sum large logarithms in the corrected coefficients in going from hard to an intermediate scale such as  $\mu_h \rightarrow \mu_i$ or a hadronic scale $\mu_i\rightarrow \mu\sim \Lambda$. In principle, one needs to separately re-sum the logarithms in the matching calculation of QCD to SCET$_I$ and then from SCET$_I$ to SCET$_{II}$. In this context, the A-type coefficients $\mathcal{C}^A$ are obtained at a hard-scale along with the soft-overlap function $\zeta_{A\perp}$. Hence, we could neglect the running of A-type matching coefficients.  The running of B-type operators requires the calculation of the one-loop anomalous dimension by keeping the UV divergent terms appearing in the SCET$_I$ loop diagrams [c.f. Fig. 3 of \cite{Hill:2004if}]. As we are dealing with axial-vector meson(s) having mass around $1.3$ GeV that lies very close to the intermediate scale ($1.5$ GeV), therefore, the resummation of SCET$_{II}$ operators is not needed in our case.

 The evolution of the B-type coefficients read as
\begin{eqnarray}
\frac{d}{dln\mu}\mathcal{C}^B _j (E,v)&=&\gamma^B _{ij}(u,v)\mathcal{C} ^B _i(E,u),\label{anomalyb}
\end{eqnarray}
where $\gamma^B$ is an anomalous dimension and it depends upon the variable $u$ that denotes the fraction of momentum carried by hard-collinear quark or gluon. The solution of the evolution equation (\ref{anomalyb}) is
\begin{eqnarray}
\mathcal{C}^{B'} _j (E,u,\mu)&=&\left(\frac{2E_F}{\mu_h}\right)^{a(\mu_h ,\mu)}e^{S(\mu_h,\mu)} \int_0 ^1 dv U_{\perp,\parallel}(u,v,\mu_h , \mu)\mathcal{C}_j ^{B'}(E,v,\mu_h) \label{CBmu},
\end{eqnarray}
where the prime notation is adopted to make it consistent with our earlier tree level coefficients in Eq. (\ref{treelev}). The functions $a(\mu_h ,\mu)$ and $S(\mu_h ,\mu)$ that appeared in Eq. (\ref{CBmu}) are defined as \cite{Bosch:2003fc,Becher:2005fg}
\begin{eqnarray}
S(\mu_h , \mu) &=& \frac{\Gamma_0}{4\beta_0 ^2}\left[\frac{4\pi}{\alpha_s (\mu_h)}\left(1-\frac{1}{r_1}-\text{ln} r_1\right)+\frac{\beta_1}{2\beta_0}\text{ln}^2 r_1 -\left(\frac{\Gamma_1}{\Gamma_0}-\frac{\beta_1}{\beta_0}(r_1 -1-\text{ln} r_1)\right)\right],\nonumber\\
a(\mu_h ,\mu)&=&-\frac{\Gamma_0}{2\beta_0}\text{ln} r_1,
\end{eqnarray}
where $\Gamma_0 =4C_F$ and
\begin{eqnarray} 
\Gamma_1 &= &4C_F \left[\left(\frac{67}{9}-\frac{\pi^2}{3}\right)C_A -\frac{20}{9}T_F n_f\right], \beta_0 =\frac{11}{3}C_A -\frac{4}{3}T_F n_F,\notag\\
\beta_1 &=& \frac{34}{3}C_A ^2 - \frac{20}{3}C_A T_F n_f -4C_F T_F n_f ,
\end{eqnarray}
along with $r_1=\alpha_s(\mu)/\alpha_s(\mu_h)$. To find the evolution function $U_{\perp,\parallel}$, we employed the method used in \cite{Hill:2004if} to solve the following RG equation using the  initial condition $U_{\perp,\parallel}(u,v,\mu_h ,\mu_h)=\delta(u-v)$
\begin{eqnarray}
\frac{d}{d \text{ln} \mu}\frac{U_{\perp,\parallel}(u,v,\mu_h ,\mu )}{\bar{u}}&=& \int_0 ^1 dy y \bar{y}^2 \frac{V_{\perp,\parallel}}{\bar{y}\bar{u}}\frac{U_{\perp,\parallel}(u,v,\mu_h ,\mu)}{\bar{y}}+w(u)\frac{U_{\perp,\parallel}(u,v,\mu_h ,\mu )}{\bar{u}}.\label{evolU}
\end{eqnarray}
\begin{figure}
  \includegraphics[width=120mm]{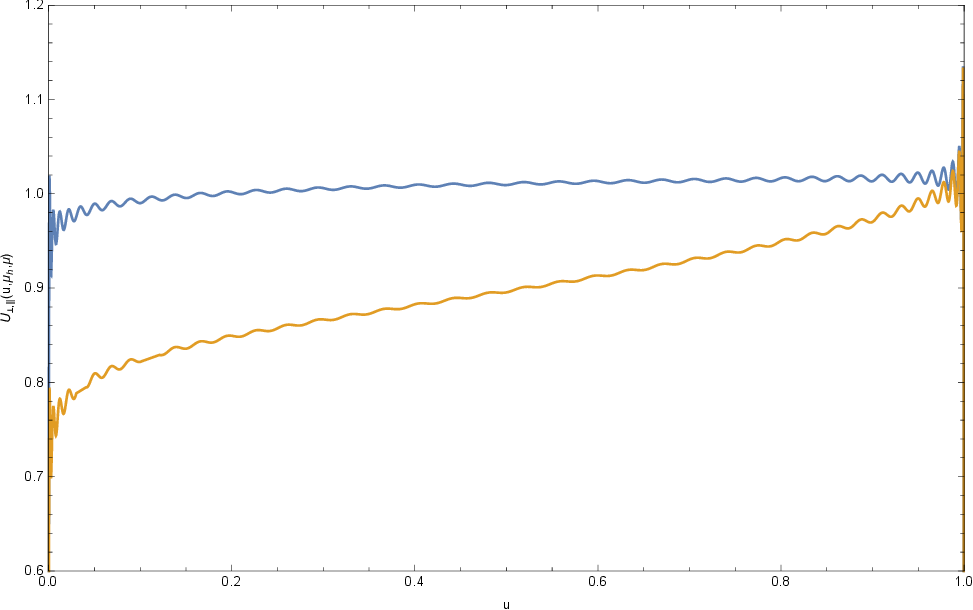}
  \caption{Evolution function with orange being $U_\perp$ and blue being $U_\parallel$ using 80 basis functions. The increase in the number of basis functions will decrease the amplitude of oscillations and smooth the curves. A slope can be fitted on these curves as a function of $u$. However, this will not affect the results calculated here.}
  \label{evolution}
\end{figure}
The functions $V_{\perp,\parallel}$ and $w(u)$ are defined in \cite{Hill:2004if} which are actually the anomalous dimension contributions. A basis function comprising of Jacobi Polynomials (obtained by solving eigenvalue equation for $V_{\perp,\parallel}$) is used to construct the solution for Eq. (\ref{evolU}). This solution is plotted in Fig. \ref{evolution} by taking a set of 80 basis functions. We can notice from the figure that the parallel evolution function $U_{\parallel}$ is almost constant at a value of $0.98$ for all values of $u$, which is represented by blue lines. The perpendicular evolution function, $U_{\perp}$, on the other hand can be parametrized as a linear function of the momentum fraction variable $u$. 

The RG-improved amplitude can now be written in a short form as
\begin{eqnarray}
\mathcal{M}=\mathcal{C}^A(E,\mu)\zeta_{A\perp}(E)&-&\frac{\sqrt{m_B}}{4}\left(\frac{2E_F}{\mu_h}\right)^{a(\mu_h ,\mu_i)}e^{S(\mu_h,\mu_i)}\int_0 ^\infty \frac{d\omega}{\omega}\phi_B (\omega, \mu)
 \int_0 ^1 du  f_{A\perp}\phi_{A_{\perp}}(u,v)\nonumber\\
&\times&\int_0^1dv \mathcal{J}_\perp \left(u,v,\text{ln}\frac{m_B \omega}{\mu^2}\right)U_{\perp}(u,v,\mu_h , \mu_i)\mathcal{C}_1 ^B(v,\mu).\label{Amp}
\end{eqnarray}

\section{Soft-overlap function and Branching Fractions}\label{branching}
The RG-improved amplitude given in Eq. (\ref{Amp}) is useful to study the branching fraction of $B\rightarrow (K_1,b_1, a_1)\gamma$ decay, which is an important observable for experimental searches. In the case of decays under consideration, we can write
\begin{eqnarray}
Br(B^+\rightarrow (K_1^+,b_1^+, h_1^+)\gamma)&= \frac{\tau_B m_B}{4\pi}\left(1+\frac{m_A ^2}{8E_F \Delta}\right)\left(1-\frac{m_A ^2}{m_B ^2}\right)|\mathcal{M}|^2 ,
\end{eqnarray}
where $m_A$ denotes the mass of the axial-vector meson. The coefficients $\mathcal{C}^A$ and $\mathcal{C}_1 ^B$ used in Eq. (\ref{Amp}) are assembled as
\begin{eqnarray}
\mathcal{C}^A&= \frac{G_F}{\sqrt{2}}V^* _{cp}V_{cb}\left[C_7 +\frac{C_F \alpha_s}{4\pi}\left(C_1 G_1(x_c)+ C_8 G_8  \right)\right]\nabla_7 C^A, \label{CA}\\
\mathcal{C}_1 ^B&= \frac{G_F}{\sqrt{2}}V^* _{cp}V_{cb}\left[C_7 +C_1 \frac{1}{3}f\left(\frac{\bar{m}_c ^2}{4\bar{u}E_F E_\gamma}\right)+C_8 \frac{\bar{u}}{3u}\right]\nabla_7 C^B _1 ,\label{CB}
\end{eqnarray}
with $p=s$ and $d$ for $K_{1}$ and $(b_1,a_1)$ mesons, respectively. To find the branching fraction, we require the soft-overlap function $\zeta_\perp(E_F)$. We could approximate this non-factorizable part by using the light-cone-sum-rules (LCSR) or SCET sum rules \cite{DeFazio:2007hw}. In this particular case, we will rely on the LCSR values of the axial-vector form factor $\mathcal{A}$, and we can write it in terms of the soft-overlap function 
\begin{equation}
\mathcal{A}(E_F)=\left(1+\frac{m_A}{m_B}\right) \frac{E_F}{\Delta}\mathcal{C}^A _{A_{1}}\zeta_A ^\perp(E_F),\label{softform}
\end{equation}
Our choice of $\mathcal{A}(E_F)$ is such that the factorizable part of $\mathcal{A}(E_F)$ is $\mathcal{O}(\alpha_s)$ suppressed compared to other form factors. This is consistent to our previous LEET results for form factors \cite{Sikandar:2019qyb}.
\begin{table}[t]
 \caption{The values of input parameters used in our numerical analysis.}\label{tab:msg2}
 \centering
    \begin{tabular}{|l|l|l|l}
    \hline
    \multicolumn{3}{|c|}{\textbf{Values}} \\ \hline
     $m_B$\qquad\qquad\qquad\quad  5.28 GeV  &$G_F$\qquad\qquad\qquad $1.16 \times 10^{-6}$ GeV&$m_{K_{1A}}$\qquad\qquad\qquad 1.31GeV      \\ \hline

     $m_{K_{1B}}$\qquad\qquad\qquad\  1.34 GeV& $m_{a_1}$\qquad\qquad\qquad\qquad 1.23GeV&$m_{b_1}$ \qquad\qquad\qquad\  1.21GeV \\ \hline

         $f_{K_{1A}\perp}$\qquad\quad 0.25$\pm$ 0.013 GeV&$f_{K_{1B}\perp}$\qquad\qquad\quad 0.19$\pm$ 0.01 GeV&  $f_{a_1}$\qquad\qquad\qquad\quad\  0.234GeV \\ \hline

  $f_{b_1\perp}$\qquad\qquad 0.18$\pm$0.008GeV&$a_{1\perp}^{(a_{1})}(1GeV)$\qquad\qquad -1.04$\pm$0.34& $a_{2\perp}^{(b_{1})}(1GeV)$\qquad\ \ \ \   0.03$\pm 0.19$\\ \hline

     $f_{B}$\qquad\qquad\qquad\qquad 0.2GeV  & $a_{0\perp}^{K_{1A}}(1GeV)$\qquad \  0.26+0.03-0.22&$a_{\perp 1}^{(K_{1A})}(1GeV)$\qquad -1.08$\pm$0.48\\ \hline

      $a_{2\perp}^{(K_{1A})}(1GeV)$\qquad\quad  0.02$\pm$0.2 & $a_{1\perp }^{(K_{1B})}(1GeV)$\quad\  $0.3+0.0-0.31$ & $a_{\perp 2}^{(K_{1B})}(1GeV)$\quad\   $-0.02\pm0.22$\\ \hline

  $\lambda_{B,+} $\qquad\qquad\qquad\ \  0.35GeV & $\tau_B$ \qquad\qquad\qquad\qquad\qquad  1.6ps &$C_1(m_b)$\qquad\qquad\qquad 1.108 \\  \hline

  $C_7^{(eff)}(m_b)$\qquad\qquad\quad -0.311  &$|V^*_{tb}V_{td}|$\quad\qquad\qquad\qquad\ \ $5\times 10^{-3}$&$|V^* _{cs}V_{cb}|$\qquad\qquad\quad $4\times 10^{-2}$\\ \hline
     $\theta_{K_1}$\qquad\qquad\qquad -34$^0\pm 13^0$&$C_8^{(eff)}(m_b)$\qquad\qquad\qquad -0.151&\\ \hline
      \end{tabular}
    \label{tab:msg2}
\end{table}
The $C^\text{A} _{A_{1}}$ is the coefficient obtained by matching A-type operators with axial-vector current $A_1 =\gamma^\mu \gamma_5$ in the full theory. It is given as
\begin{equation}
C^\text{A} _{A_{1}}= 1-\frac{\alpha_s C_F}{4\pi}\left[\left(\frac{1}{1-x}-3\right)\text{ln}(x) +2\text{ln}^2(x) +2 \text{ln}^2\left(\frac{\mu_i}{m_b} \right)^2 + 2\text{Li}_2(1-x)-(4\text{ln}(x)-5)\text{ln}\left(\frac{\mu_i}{m_b}\right)+\frac{\pi^2}{12}+6\right],\label{CV}
\end{equation}
with $x=2E_F/m_b$. As physical $K_1$ meson is a mixture of $^1 P_1 (K_{1B})$ and $^3 P_1 (K_{1A})$ states therefore, their corresponding form factors will mix too. The two physical mass states $K_{1}(1270)$ and $K_{1}(1400)$ depend upon the mixing angle $\theta_{K_1}$ and the corresponding form-factors are given by
\begin{eqnarray}
\mathcal{A}^{(1270)}(E_F)&=&\frac{m_B +m_{K{_{1}(1270)}}}{m_B +m_{K{_{1A}}}}\mathcal{A}^{K_{1A}}(E_F)\text{sin}\theta_{K_{1}}+\frac{m_B +m_{K{_{1}(1270)}}}{m_B +m_K{_{1B}}}\mathcal{A}^{K_{1B}}(E_F)\text{cos}\theta_{K_{1}},\nonumber\\
\mathcal{A}^{(1400)}(E_F)&=&\frac{m_B +m_{K{_1(1400)}}}{m_B +m_{K{_{1A}}}}\mathcal{A}^{K_{1A}}(E_F)\text{cos}\theta_{K_{1}}-\frac{m_B +m_{K{_1(1400)}}}{m_B +m_{K{_{1B}}}}\mathcal{A}^{K_{1B}}(E_F)\text{sin}\theta_{K_{1}}.\label{formix}
\end{eqnarray}
These form-factors are given as a function of final state meson's energy with maximum recoil, i.e., $q^2=0$. The LCDAs for final state axial-vector mesons at twist-2 for $1^1P_1$ $(K_{1B},b_1)$ and $1^3P_1$ $(K_{1B},a_1)$ states are calculated in \cite{Yang:2008xw} and these are given as
\begin{eqnarray}
\Phi_\perp^{1 ^1 \text{P}_1}(u) &=& 6u\bar{u}\left[1+3a_1^\perp\Upsilon +a_2^\perp\frac{3}{2}(5\Upsilon^2 -1)\right],\nonumber\\
\Phi_{\perp}^{1 ^3 \text{P}_1}(u)&=&6u\bar{u}\left[a_0^\perp +3a_1^\perp\Upsilon +a_2^\perp\frac{3}{2}(5\Upsilon^2 -1)\right],
\end{eqnarray}
with $\Upsilon = 2u-1$. The  LCDA for $1 ^1\text{P}_1$ and $1 ^3\text{P}_1$ states are normalized as $\int_0 ^1 du \phi_\perp(u)/u =1$ and $\int_0 ^1 du \phi_\perp(u)/u =a_0^\perp$, respectively. The Gagenbaur moments and decay constants at a scale $\mu=1$GeV are given in Table \ref{tab:msg2} . Ideally, one needs to RG run these parameters from intermediate scale to lower scale - but RG-improvement of these hadronic parameters to find branching fractions is not helpful as the uncertainties associated with their values at present are large. For physical $K_1$ meson states these LCDAs mixes just like the form factors given in Eq. (\ref{formix}) and these can be written as \cite{Yang:2007zt}
\begin{eqnarray}
\Phi^{K_1 (1270)}_\perp (u)&=&\frac{f^{K_{1A}}_\perp}{f^{K_{1}(1270)}_\perp}\text{sin}\theta_{K_1}\Phi^{K_{1A}}_ \perp(u) +\frac{f^{K_{1B}}_\perp}{f^{K_{1}(1270)}_\perp}\text{cos}\theta_{K_1}\Phi^{K_{1B}}_ \perp(u),\nonumber\\
\Phi^{K_1 (1400)}_\perp (u)&=&\frac{f^{K_{1A}}_\perp}{f^{K_{1}(1400)}_\perp}\text{cos}\theta_{K_1}\Phi^{K_{1A}}_ \perp(u) -\frac{f^{K_{1B}}_\perp}{f^{K_{1}(1400)}_\perp}\text{sin}\theta_{K_1}\Phi^{K_{1B}}_ \perp(u).
\end{eqnarray}
Many different models deals with the $B-$meson distribution amplitudes see e.g. \cite{Braun:2003wx}. Recently, the BABAR experiment \cite{Aubert:2009ya} has constrained the value of $\lambda_B$ in range $0.30 - 0.67$ GeV. In ref. \cite{Beneke:2011nf}, Beneke et al. have further improved this range by considering power- and the radiative corrections. In the present analysis, we used the optimal value $\lambda_B \sim 0.35$GeV. The decay constants for the mesons are taken from \cite{Yang:2008xw}.

 From Eq. (\ref{softform}), by taking the values of  $\mathcal{A}_{K_{1A}}(0)= 0.45\pm 0.08$ and $\mathcal{A}_{K_{1B}}(0)= -0.37\pm 0.06$ for $K_1 (1270)$ and $K_1(1400)$ mixed $K_1$ states and $\mathcal{A}_{a_1}(0)=0.48\pm 0.08$, $\mathcal{A}_{b_1}(0)=-0.025\pm 0.05$, the values of soft form factor $\zeta_{\perp}(E_F)$ for different axial-vector states are obtained to be
 \cite{Yang:2008xw}
\begin{eqnarray}
\zeta^{K_1(1270)}_{\perp}(E_F)&=&-0.47\pm 0.03,\notag\\
\zeta^{K_1(1400)}_{\perp}(E_F)&=&0.14 \pm 0.06,\notag\\
\zeta^{a_1(1260)}_{\perp}(E_F)&=&0.37\pm 0.08,\notag\\
\zeta^{b_1(1235)}_{\perp}(E_F)&=&-0.22\pm 0.04, \label{zetavalues}
\end{eqnarray}
where we took $\theta_{K_1}=-34^{o}$, and the uncertainties are associated to the LCSR value of the axial-vector form factor $\mathcal{A}$ for different axial-vector mesons.
It is important to emphasis that the value of $\zeta_{K_1}$ for $K_{1}(1400)$ is highly sensitive to the value of the mixing angle. We can see that its central value lies in the range $[0.03,0.24]$ when we vary $\theta_{K_1}\in[-46^{o},-22^{o}]$, i.e. $\mp12^{o}$ around $\theta_{K_1}=-34^{o}$. However, the values of the form factor for $K_{1}(1270)$ do not change significantly and they lie in the range $[-0.49,-0.45]$. 

In Eq. (\ref{zetavalues}), we can see that the value of $\zeta_{K_1(1400)}$ is almost three times smaller than $\zeta_{K_1(1200)}$, and this suppression of the form factor due to the mixing angle for the physical state $K_1(1400)$ makes the convolution integral almost equally important. Also, the hadronic uncertainties arising due to the spectator part affect the branching ratio of $K_{1}(1400)$ more compared to $K_1 (1270)$. The LCSR value $\mathcal{A}(0)=0.45$ for $a_1$ is significantly different from the one found in \cite{Li:2009tx} i.e. $\mathcal{A}(0)=0.26$. Hence, to have a reliable estimate of the branching fraction for the radiative $B \to a_1\gamma$ decay, we desire to have a more accurate value of $\mathcal{A}(0)$. To do so, one would probably have to rely on SCET sum rules similar to \cite{DeFazio:2007hw,Gao:2019lta} or use lattice QCD results to fix the overlap function. The other option for its fixing is to use an experimentally measured branching fraction when available. Nevertheless, these branching fractions calculated here and tabulated in Table \ref{tab:BF} are within a measurable range of LHCb and the future $B$-factories.

 We compared the theoretical values of the branching fractions for $B\to K_{1}(1270,1400)\gamma$ decay obtained here with the corresponding results obtained using ADS/QFT approach \cite{Momeni:2018udf}, LCQCD Sum Rules \cite{Hatanaka:2008xj} and the experimental values in Table\ref{tab:BF}. The uncertainty in branching fraction for our work in Table \ref{tab:BF} is due to LCSR. The result for $K_1(1270)$ is less impacted by hadronic uncertainties and is well within the range of observed experimental value.

Keeping in view that the branching ratio of $B \to K_{1}(1270)\gamma$ is already measured experimentally, it will be useful to give the following ratios 
\begin{eqnarray}
\mathcal{R}[\mathcal{B}(B\rightarrow K_1(1270)\gamma)/(\mathcal{B}(B\rightarrow a_1(1260)\gamma)]&=&68.6,\nonumber\\
\mathcal{R}[\mathcal{B}(B\rightarrow K_1(1270)\gamma)(\mathcal{B}(B\rightarrow b_1(1235)\gamma)]&=&178.
\end{eqnarray}
\begin{table}[t]
\caption{The calculated values of the branching fractions of $B\to \left(b_1,\; a_{1},\; K_{1}(1270,1400)\right)\gamma$ decays. Using $\theta_{K_1} = -34^{o}$ the results of the branching fractions of $B\to \left(K_{1}(1270,1400)\right)\gamma$ calculated in this work are compared with ADS/QFT  \cite{Momeni:2018udf}, LCQCD Sum Rules \cite{Hatanaka:2008xj}, and the experimental values \cite{PDG:2020zy}. All the results are of the order $10^{-6}$.}\label{tab:BF} 
\centering 
\begin{tabular}{|c|c|c|c|c|} 
\hline\hline 
Decay Channel & ADS/QFT Approach \cite{Momeni:2018udf} & LCQCD Sum Rules \cite{Hatanaka:2008xj}& This work & Exp. values \cite{PDG:2020zy} \\ [0.5ex] 
\hline
$K_1^+ (1270)\gamma$ & $71\pm 23 $& $66^{+21+30+2+6}_{-12-24-4-12}$  & $48 \pm 6$ & $44^{+7}_{-6}$\\ \hline
$K_1^+ (1400)\gamma$& $15.6\pm 10.4$ &$6.5^{+4+2.6+0.1+11.9}_{-2.2-0-0.2-5.9}$&$5.4\pm4.8$  &$10^{+5}_ {-4}$\\ \hline
$a_1^+ (1260)\gamma$&--&--&$0.7\pm 0.037$&--\\ \hline
$b_1^+ (1235)\gamma$&--&--&$0.27\pm 0.04$&--\\ \hline

 \end{tabular}

\end{table}
It is tempting to find the branching fractions for other $1 ^1P_1$ states $(h_1(1170),\;h_1(1380))$ and $1 ^3P_1$ states $(f_1(1285),\;f_1(1420))$. These mass states arise due to the mixing of the singlet and octet states, e.g.,
\begin{eqnarray}
|h_1(1170)\rangle&=|h_1 \rangle \text{cos}\theta_{1 ^1P_1}+|h_8 \rangle \text{sin}\theta_{1 ^1P_1},\nonumber\\
|h_1(1380)\rangle&=-|h_1 \rangle \text{sin}\theta_{1 ^1P_1}+|h_8 \rangle \text{sin}\theta_{1 ^1P_1}.
\end{eqnarray}
Similarly, the $f_1$ states mixes with angle $\theta_{1 ^3P_1}$. Previously, the singlet states showed the anomalous behavior in their masses, and branching fraction, e.g., the branching fraction of $B\rightarrow K^+ \eta^\prime$ is almost six times larger than that of $B\rightarrow K^+\eta$. For radiative $\phi$ decay, the branching ratio $\mathcal{R}[\phi\rightarrow f_0\gamma/\phi\rightarrow a_0\gamma]\sim 6.1$ \cite{PDG:2020zy}. We also expect such an enhancement in the branching fraction of the singlet states in radiative $B-$decays.
To take into account the singlet-amplitude, a new set of operators can be introduced at leading power to account, e.g., from Fig. 7 of \cite{Becher:2005fg}. This would be like $b\rightarrow sq\bar{q}$ where $q\bar{q}$ hadronizes to $(|f_1\rangle ,|h_1\rangle)$ states. For the partons that carry small momentum, there arise soft-collinear modes that interact with the collinear and soft parts at leading order. These contributions appear as end-point singularities and spoil our factorization. Therefore, the estimated values of the branching fractions for singlet-states would not be precise enough. It may be possible to improve this estimate by introducing a new non-perturbative parameter in the factorization relation, which is on the line of QCD factorization scheme presented in \cite{Beneke:2002jn}.  However, it lies beyond the scope of the present study. 

 \section{Summary} \label{summary}

The radiative $B-$ to axial-vector meson decays are studied at next-to-leading order in $\alpha_s$ and at the leading-power of $1/m_b$ in the SCET. The effective theory approach and power counting helped us in collecting diagrams that contribute to the matching calculations. These diagrams are matched onto SCET$_I$ operators at one-loop order for A-type operators and the tree level for the B-type operators. The diagrams with photon emission from the spectator anti-quark contribute to leading power - but vanish when the matrix elements are calculated for the transversely polarized axial-vector meson. The large perturbative logarithms are resummed for the case of B-type operators using one-loop SCET$_I$ diagrams. Estimating the soft-overlap functions a.k.a. soft form factors from LCSR, the branching fractions of $B\to(K_1(1270), K_1 (1400), b_1 (1235), a_1 (1260))\gamma$ decays are calculated. We find that the value of the form factor for $K_1 (1400)$ depends greatly on the mixing angle, and so does the corresponding branching fraction. The results for both physical $K_1$ mesons are within the measured experimental uncertainty. For $(a_1 , b_1)$ mesons the values lie within the reach of LHCb and the future B-factories. Hence, we hope that the future developments in the experimental and theoretical side will help us in having a precise estimate of the values of the soft overlap functions of $b_1$ and $a_1$ mesons.

\section{Appendix}\label{app}
For the matching of $Q_1$ and $Q_8$ on A-type operators, the functions $G_1$ and $G_8$ were required. These are taken from \cite{Bosch:2001gv}, and their explicit form is
\begin{eqnarray}
G_1(x)&=&\frac{-104}{27}\text{ln}\frac{\mu }{m_b}-\frac{833}{162}-\frac{20 \pi i}{27}+8\frac{8\pi^2}{9}x^{3/2} \nonumber\\
&+&\frac{2}{9}\left[48 +30 \pi i-5\pi^2-36\xi(3)+(36+6\pi i-9\pi^2)\text{ln}(x)\right.\nonumber\\
&&\left. \qquad\qquad +(3+6\pi i)\left(\text{ln}(x)\right)^2 +\left(\text{ln}(x)\right)^3\right]x\nonumber\\
&+&\frac{2}{9}\left[18 +2\pi^2 -2\pi^3 +(12-6\pi^2)\text{ln}x +6\pi i \left(\text{ln}(x)\right)^2+\left(\text{ln}(x)\right)^3 \right]x^2\nonumber\\
&+&\frac{1}{27}\left[-9+112\pi i-14\pi^2+(182-48\pi i)\text{ln}(x) -126\left(\text{ln}(x)\right)^2 \right]x^3 +\mathcal{O}(x^4),\nonumber\\
G_8&=& \frac{8}{3}\text{ln} \left(\frac{\mu }{m_b}\right)+\frac{11}{3}+\frac{2\pi i}{3}-\frac{2\pi^2}{9},
\end{eqnarray}
here $x=m_c^2/m_b^2$. The function $f(z)$ where $z=\frac{m_q^2}{4EE_\gamma \bar{u}}$ given in the matching of $Q_1$ operators to B and C type currents is given as
\begin{eqnarray}
f(z)=\left\{
  \begin{array}{@{}ll@{}}
    1+4z\left(\text{arctanh}(\sqrt{1-4z})-i\frac{\pi}{2}\right)^2, & \text{for}\ z<1/4 \\\\

    1-4z\ \text{arctan}^2\frac{1}{\sqrt{4z-1}}, & \text{for}\ z>1/4.
  \end{array}\right.
 \end{eqnarray}

\end{document}